\documentclass[10pt]{iopart}
\expandafter\let\csname equation*\endcsname\relax
\expandafter\let\csname endequation*\endcsname\relax

\usepackage{amsmath}
\usepackage{graphicx}
\usepackage{lipsum}
\usepackage{bm}
\usepackage{tabu}
\usepackage[utf8]{inputenc}
\usepackage[T1]{fontenc}
\usepackage{mathptmx}
\usepackage{color}
\usepackage{xcolor}
\usepackage{cite}
\usepackage{iopams}

\graphicspath{{images/}}   
\usepackage{chapterbib}
\usepackage{comment}

\newcommand{\vect}[1]{\ensuremath{\mathbf #1}}
\newcommand{\hatt}[1]{\ensuremath\hat{{\mathbf #1}}}

\begin{document}
\title[Highly-stable generation of vector beams through a common-path interferometer and a DMD]{Highly-stable generation of vector beams through a common-path interferometer and a DMD}

\author{Benjamin Perez-Garcia$^1$, Francisco Israel Mecillas Hern\'andez$^2$ and Carmelo Rosales-Guzm\'an$^{2,3}$}

\address{$^1$ Photonics and Mathematical Optics Group, Tecnologico de Monterrey, Monterrey Mexico}
\address{$^2$ Centro de Investigaciones en Óptica, A.C., Loma del Bosque 115, Colonia Lomas del campestre, C.P. 37150 León, Guanajuato, Mexico}
\address{$^3$ Wang Da-Heng Collaborative Innovation Center, Heilongjiang Provincial Key Laboratory of Quantum Manipulation and Control, Harbin University of Science and Technology, Harbin 150080, China.}

\ead{b.pegar@tec.mx}

\begin{abstract}
Complex vector modes of light, non-separable in their spatial and polarisation degrees of freedom, are revolutionising a wide variety of research fields. It is therefore not surprising that the generation techniques have evolved quite dramatically since their inception. At present it is common to use computer-controlled devices, among which digital micromirror devices have become popular. Some of the reason for this are their low-cost, their polarisation-insensitive and their high-refresh rates. As such, in this manuscript we put forward a novel technique characterised by its high stability, which is achieved through a common-path interferometer. We demonstrate the capabilities of this technique experimentally, first by generating arbitrary vector modes on a higher-order Poincar\'e sphere, secondly, by generating vector modes in different coordinates systems and finally, by generating various vector modes simultaneously. Our technique will find applications in fields such as optical manipulations, optical communications, optical metrology, among others. 
\end{abstract}

\ioptwocol
\maketitle
\section{Introduction}

The on-demand manipulation of the properties of light, such as, amplitude, phase, polarisation or frequency, is amongst the most popular topics of late \cite{roadmap}. In particular, complex vector modes, featuring a non-homogeneus transverse polarisation distribution, are attracting the attention of a wide research community due to their potential applications in fields such as, optical manipulations, high-resolution microscopy, optical metrology, classical and quantum communications, amongst others \cite{rosales2018review,Zhan2009,hu2019situ,toppel2014classical,li2016high,Yuanjietweezers2021,Ndagano2017,bhebhe2018,zhao2015high}. Such vector modes are mathematically described as a non-separable superposition of the spatial and polarisation Degrees of Freedom (DoFs) in a similar way to entangled states, a feature that has given rise to a wide variety of quantum-like analogies \cite{McLaren2015,Ndagano2016,Zhaobo2020,Aiello2015,konrad2019quantum,Ndagano2017,Qian2015,Bhebhe2018a}. In regards to the generation techniques, they have evolved quite dramatically since their inception, specially after the invention of liquid crystal Spatial Light Modulators (SLMs), which enabled for the first time the on-demand generation of vector beams with arbitrary spatial shapes and polarisation distributions \cite{Maurer2007,Rosales2017,Moreno2012,Mitchell2017,Rong2014,SPIEbook}. More recently, Digital Micromirror Devices (DMD) gain popularity due to their high-speed refresh rates, polarisation-insensitive properties, as well as their low cost \cite{Hu2018,Ren2015,Chen2015DMD,Mitchell2016,Goorden2014,Lerner2012,Hu2021Random,Scholes2019,Hu2022,Rosales2020,Gong2014}. Their potential has been demonstrated in the generation of vector modes in different coordinates systems \cite{Liyao2020,Zhao2022,Rosales2021,Hu2021,hu2021generation}. Since the first DMD generation techniques implemented without taking advantage of their polarisation-insensitive properties (see for example \cite{Mitchell2016}), techniques have evolved in various ways. This is the case of the technique reported in \cite{Rosales2020}, where the polarisation-insensitive property of DMDs was exploited for the first time. Here, a DMD is impinged with two expanded beams at opposite angles, 2$^\circ$ and -2$^\circ$ respectively, bearing orthogonal polarisation. The DMD is then addressed with a multiplexed hologram encoding the two holograms that constitute the vector beam, each with a unique spatial grating to deflect the first diffraction order of each beam along the same propagation axis, where the vector beam is generated. In this approach it is crucial to place the DMD exactly at the plane where the two input beams cross each other. 

In this manuscript we propose a novel generation technique which takes full advantage of the properties of DMDs and the high-stability feature of a common-path interferometric array. Even though a similar technique with SLMs has been previously demonstrated, this is limited to a maximum refresh rate of 60 Hz and to an input horizontal polarisation due to the inherent properties of SLMs \cite{Perez-Garcia2017}. On the contrary, our present technique does not require a specific input polarisation, even though, diagonal polarisation might be preferred. In addition, the DMD screen is not split into two sections but rather it incorporates a multiplexing approach to generate both constituting beams from a single hologram, which allows an easier alignment. Crucially, this technique also allows the simultaneous generation of multiple vector beams with independent properties, as we also demonstrated experimentally \cite{Rosales2017}.
    
\section{Theoretical considerations}
\subsection{Tailoring light with a digital micromirror device}

\begin{figure}[tb]
    \centering 
    \includegraphics[width=0.48\textwidth]{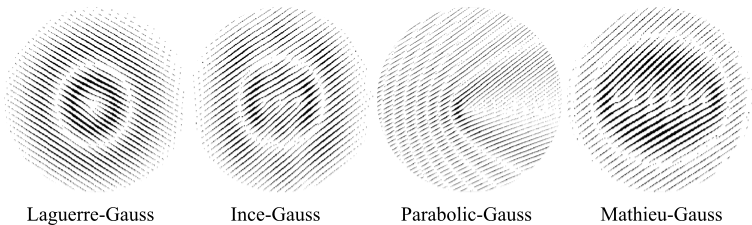}
    \caption{Example of the binary holograms encoded on a DMD to generate, from left to right, the modes $L G_{0}^{1}$, $IG_{5}^{3}$, $PG$ and , $MG_{6}^{6}$ }
    \label{holograms}
\end{figure}

As it has been widely demonstrated, DMDs have the capability to modulate the amplitude $A(x, y)$ and phase $\phi(x,y)$ of a complex optical field $U(x, y)=A(x, y) e^{i \phi(x, y)}$ through a binary hologram \cite{Lee79}. The transmittance function that allows to encode the amplitude and phase information, has the specific form \cite{Mitchell2016}
\begin{equation} \label{transmittance_function}
  T(x, y)=\frac{1}{2}+\frac{1}{2} \operatorname{sgn}\{\cos [p(x, y)]+\cos [q(x, y)]\},
\end{equation}
where the information of the amplitude and phase are encoded in the terms $q(x, y)$ and $p(x, y)$, respectively, and computed through the expressions
\begin{equation}
\nonumber
\begin{array}{l}
q(x, y)=\arcsin \left[A(x, y) / A_{\max }\right],\\
p(x, y)=\phi(x, y)+2 \pi(v x+\eta y).
\end{array}
\end{equation}
Here, $\operatorname{sgn}\{\cdot\}$ is the sign function and $A_{max}$ represents the maximum value of the amplitude $A(x,y)$. Further, the second term in $p(x,y)$ is an additional linear phase grating that controls the separation of the multiple diffraction orders according to the frequency specified by the parameters $\nu$ and $\eta$. As an example, Fig. \ref{holograms} shows the binary holograms that are displayed on a DMD to generate, from left to right, a Laguerre- (LG), an Helical Ince- (HIG), a Travelling Parabolic- (TPG) and a Helical Mathieu-Gauss (HMG) mode. The frequencies of the linear phase gratings where chosen for display purposes and do not necessary correspond to values used in the experiments. 

In regards to the generation of complex vector modes, we rely on their mathematical expression, which is a non-separable weighted superposition of the spatial and polarisation DoFs. Mathematically, such superposition (at the plane $z=0$) is represented as \cite{Galvez2015,galvez2012poincare}
\begin{equation}
    \mathbf{U}(x, y)=\cos \theta\:\: U_{R}(x, y) \hatt{e}_{R}+e^{i 2\alpha} \sin \theta\:\: U_{L}(x, y) \hatt{e}_{L}
    \label{vector_bea},
\end{equation}
where $U_R$ and $U_L$ are two orthogonal scalar modes endowed with orthogonal polarisation, right-- and left--handed, respectively.  Further, $\hatt{e}_{R}$ and $\hatt{e}_{L}$ represent the unitary vectors associated to right-- and left--handed circular polarisation.  The contribution of each scalar field $U_R(x, y)$ and $U_L(x, y)$ is controlled by the parameter $\theta\in[0,\pi/2]$. Finally, an additional inter-modal phase $\alpha\in[0,\pi]$ allows to further control the polarisation distribution of the vector mode along the transverse plane. Crucially, the scalar fields $U_R(x, y)$ and $U_L(x, y)$ can take the specific form of any of the orthogonal sets of solutions of the wave equation, in its exact or paraxial form, in the different coordinate systems \cite{Liyao2020,Zhao2022,Rosales2021,Hu2021,hu2021generation}.

\section{Generation of highly-stable arbitrary vector modes}
\subsection{Experimental setup}
\begin{figure}[tb]
    \centering 
    \includegraphics[width=0.48\textwidth]{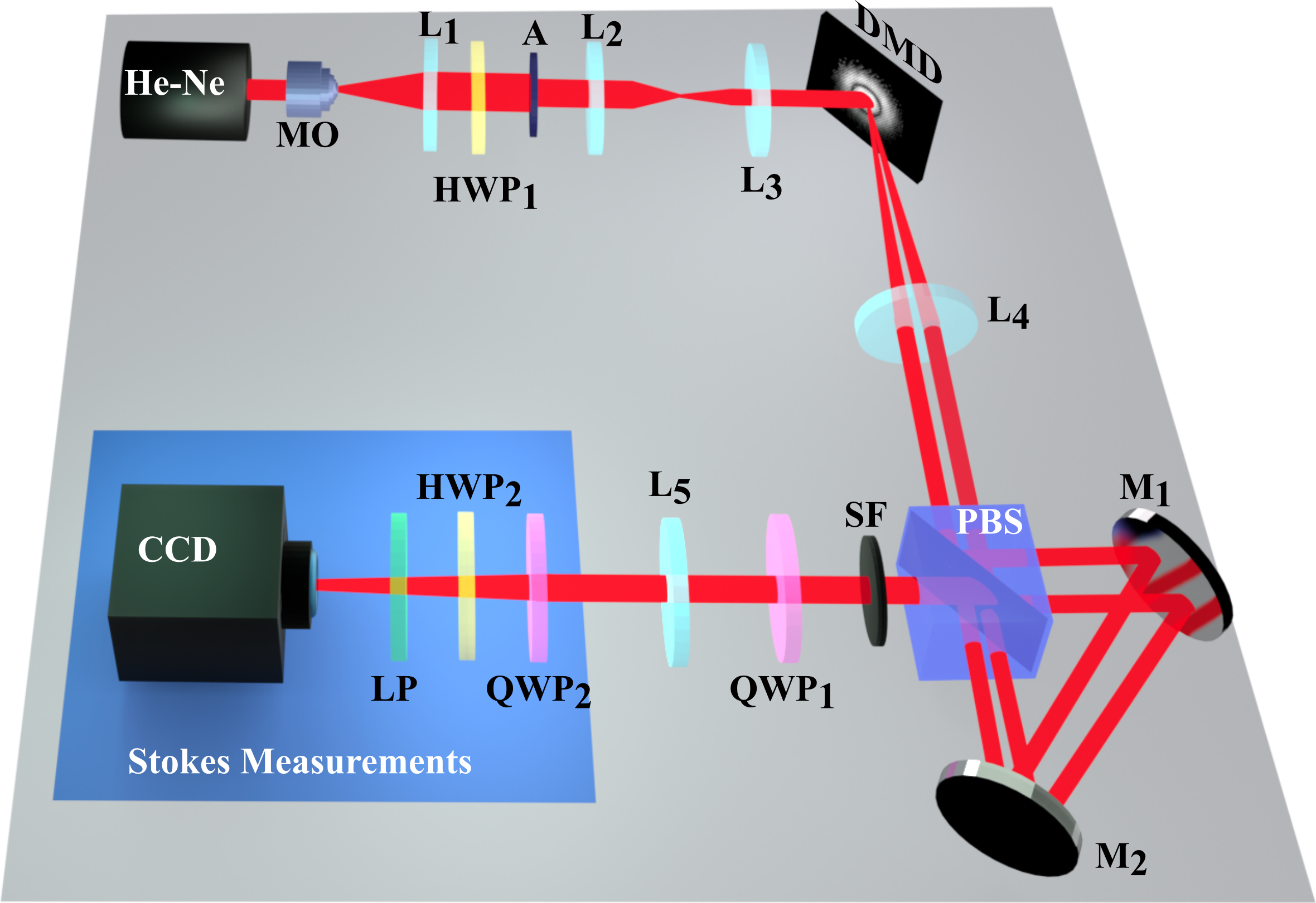}
    \caption{Schematic representation of our highly stable device. An expanded and collimated He-Ne laser with diagonal polarisation impinges onto a DMD where the two binary holograms required to generate a complex vector mode is displayed. The first diffraction order of the two output modes are sent to a common-path interferometer for the generation of vector modes in the linear polarisation basis, which are switched to the circular basis using a Quarter Wave Plate (QWP$_1$). The vector modes are analysed through stokes polarimetry using QWP$_2$, HWP$_2$, a linear polariser (LP) and a CCD camera. MO: Microscope Objective, A: aperture; HWP: Half-Wave Plate, QWP: Quarter-Wave Plate, LP: Linear Polarizer, PBS: Polarising Beam Splitter, L: Lens, M: Mirror, SF:  Spatial Filter,  CCD: Charge-Coupled Device Camera.}
\label{setup_experiment}
\end{figure}
The novel experimental technique to generate vector modes is based on a polarisation-insensitive Digital Micromirror Device (DMD, DLCR4710EVM-G2 with pixel size 5.4 $\mu$m from Texas Instruments) and a common-path interferometer of the Sagnac type, as schematically illustrated in figure \ref{setup_experiment}. Here, a continuous wave laser beam ($\lambda$ = 633 nm) with horizontal polarisation, expanded and collimated by a microscope objective MO and lens L$_1$ ($f=300$ mm) to approximate a flat wavefront, is sent through a half-wave plate oriented at $22.5^\circ$ (HWP$_1$) and rotates the polarization state to +45$^\circ$. Lenses L$_2$ ($f =$ 300 mm) and L$_3$ ($f =$ 200 mm) form a telescope to image the flat wavefront onto the DMD. In the DMD screen, two multiplexed binary holograms, which contain the amplitude and phase information of the constituting orthogonal modes, are displayed. In each hologram, an independent linear phase grating is added to separate each orthogonal mode along different propagation directions. More precisely, the spatial frequency of each grating has the same value in the $y$ direction and apposite in the $x$ direction. In this way, when the flat wavefront impinges onto the DMD, two orthogonal modes with diagonal polarisation travelling along different paths are generated in the first diffraction order, which are isolated from higher diffraction orders by the spatial filter (SF). Both orthogonal modes, with diagonal polarisation, are fed to a Sagnac \cite{hariharan2003optical} interferometer constituted by a polarising beam splitter (PBS) and two mirrors (M$_1$ and M$_2$).  The PBS splits both beams into their horizontal and vertical polarisation components, which travel along a common optical path but in opposite directions. The orientation of the mirrors in combination with the DMD are adjusted to ensure the overlap of the orthogonal spatial modes that carry orthogonal polarisation components, horizontal and vertical, respectively, the other two modes are spatially filtered with the help of SF.  In this way, at the output end of the interferometer, a vector beam in the linear polarisation basis is generated.  Then, a quarter-wave plate (QWP$_1$) with its fast axis orientated at 45$^\circ$ is placed to change the vector beam from the linear to the circular polarisation basis.  Lenses L$_4$ and L$_5$ ($f=300$ mm), form a 4f optical system between the DMD and the charged coupled device (CCD DCX Thorlabs, 4.65 $\mu$m pixel size), which is used to observe and to characterise the vector modes. The characterisation is performed via a reconstruction of their transverse polarisation distribution as well as through  a measure of their concurrence, both implemented through Stokes polarimetry \cite{Goldstein2011,Selyem2019,Zhao2019}.

\subsection{Polarisation reconstruction and concurrence}

\begin{figure}[tb]
    \centering 
    \includegraphics[width=0.5\textwidth]{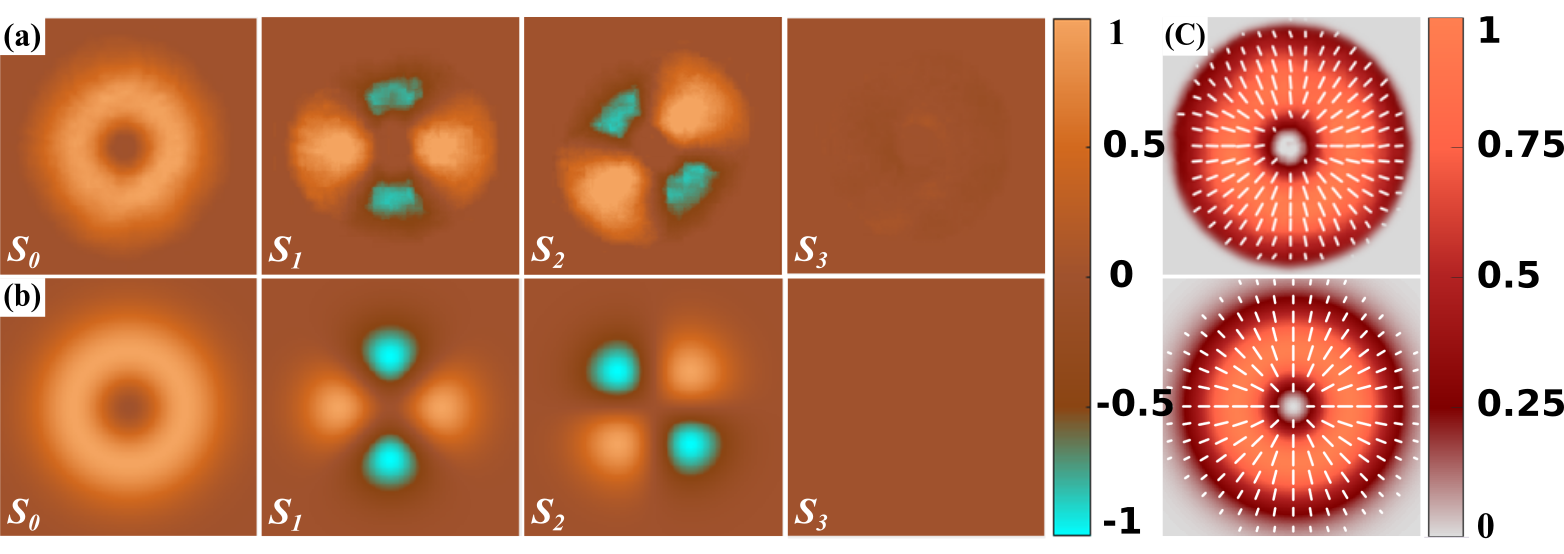}
    \caption{Experimental (a) and theoretical (b) Stokes parameters $S_0$, $S_1$, $S_2$ and $S_3$ used to reconstruct the polarisation distribution of the vector mode $\vect{U}^{1,0}_{LG}$, with $(\theta,\alpha)=(\pi/4,0)$ shown in (c).}
    \label{stokes_reconstruction}
\end{figure}

To reconstruct the transverse polarisation distribution, the Stokes parameter are determined from four intensity measurements according to the equations \cite{Goldstein2011}
\begin{equation} \label{stokes_parameters}
    \begin{array}{ll}
S_{0}=I_{0}, & S_{1}=2 I_{H}-S_{0}, \\
S_{2}=2 I_{D}-S_{0}, & S_{3}=2 I_{R}-S_{0},
\end{array}
\end{equation}
where  $I_R$, $I_L$, $I_H$, $I_D$  represent the intensities associated to right and left circular polarisation, horizontal and diagonal polarisation, respectively, with the total intensity represented by $I_0$. Such intensities are measured with the help of QWP, a HWP and a LP, as schematically shown in Fig.\ \ref{setup_experiment} and recorded with the CCD camera (Stokes measurement stage). More precisely, the intensities associated to horizontal and diagonal polarisation are measured using a HWP$_2$ with its fast axis orientated at 0$^\circ$ and 22.5$^\circ$, respectively, followed by a linear polariser orientated at 0$^\circ$.The intensities associated to the circular and left polarisation are measured using QWP$_2$ with its fast axis at -45$^\circ$ and 45$^\circ$, respectively, followed by a linear polariser orientated at 0$^\circ$. An example of the experimental Stokes parameters reconstructed from such intensity measurements, are shown in Fig.\ \ref{stokes_reconstruction} (a), compared with a numerical simulation Fig. \ref{stokes_reconstruction} (b), for the specific case of the radially polarised vector vortex mode.  A cylindrical vector vortex beam can be expressed as
\begin{align}\label{eq:CVB}
    \vect{U}^{\ell,p}_{LG}(\rho, \varphi)=\cos \theta\:\: LG^{\ell}_{p}(\rho,\varphi)\hatt{e}_{R} + e^{i 2\alpha} \sin \theta\:\: LG^{-\ell}_{p}(\rho,\varphi)\hatt{e}_{L},
\end{align}
where $LG_p^\ell(\rho,\varphi) \propto (\rho/w_0)^{|\ell|}\: e^{-\rho^2/w_o^2}\: \mathcal{L}_p^{\ell}(2\rho^2/w_0^2) e^{i\ell\varphi}$ is the Laguerre-Gauss beam at $z=0$ \cite{Forbes2014book}.  $(\rho,\varphi)$ are the polar coordinates, $w_0$ is the beam waist, $\ell$ is know as the topological charge, $p$ is the radial index and $\mathcal{L}_p^{\ell}(\cdot)$ is the associated Laguerre polynomial.  For our particular case, we used the parameters $\theta$ = $\frac{\pi}{4}$, $\alpha = 0$, $\ell = 1$ and $p = 0$. The reconstructed polarisation for such case is shown in Fig. \ref{stokes_reconstruction} (c), experiment on the top, numerical simulation on the bottom.

The quality of the generated vector beams, was quantified through the degree of concurrence $C$ defined as \cite{Selyem2019}
 \begin{equation}
 \label{concurrence}
     C=\sqrt{1-\left(\frac{\mathbb{S}_{1}}{\mathbb{S}_{0}}\right)^{2}-\left(\frac{\mathbb{S}_{2}}{\mathbb{S}_{0}}\right)^{2}-\left(\frac{\mathbb{S}_{3}}{\mathbb{S}_{0}}\right)^{2}},
 \end{equation}
with $C\in[0,1]$ where 0 is assigned to pure scalar modes and 1 to pure vector modes. Here $\mathbb{S}_{i}=\iint_{-\infty}^{\infty} S_{i} d A$ is the integral of the Stokes parameters over the entire transverse plane.
 
\begin{figure}[tb]
    \centering 
    \includegraphics[width=0.48\textwidth]{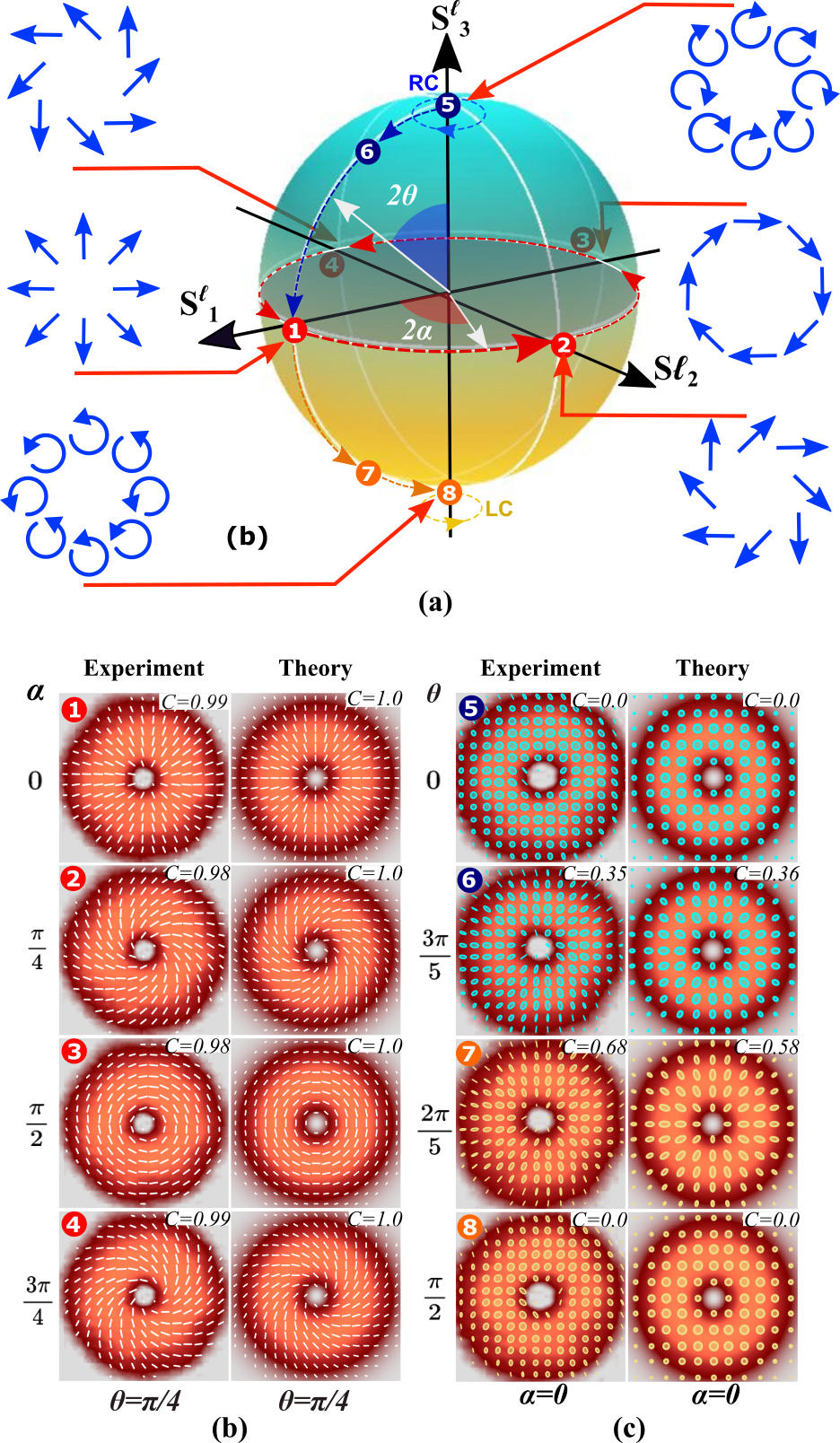}
      \caption{ (a) Geometric representation of $\vect{U}^{\ell,p}_{LG}$ on the HOPS with the numbers representing the locations of exemplary pure (red dashed line) and non-pure (blue and orange dashed lines) vector modes. The reconstructed transverse polarisation distribution overlapped with the intensity profile are shown in (b)  and (c), respectively, experiment on the left column and numerical simulations on the right. The specific coordinates $(2\theta, 2\alpha)$ are shown on each figure along with their corresponding $C$ value shown on the top-left corner. }
    \label{Poincare_Sphere}
\end{figure}

\subsection{Experimental results}

  \begin{figure*}[tb]
    \includegraphics[width=.98\textwidth]{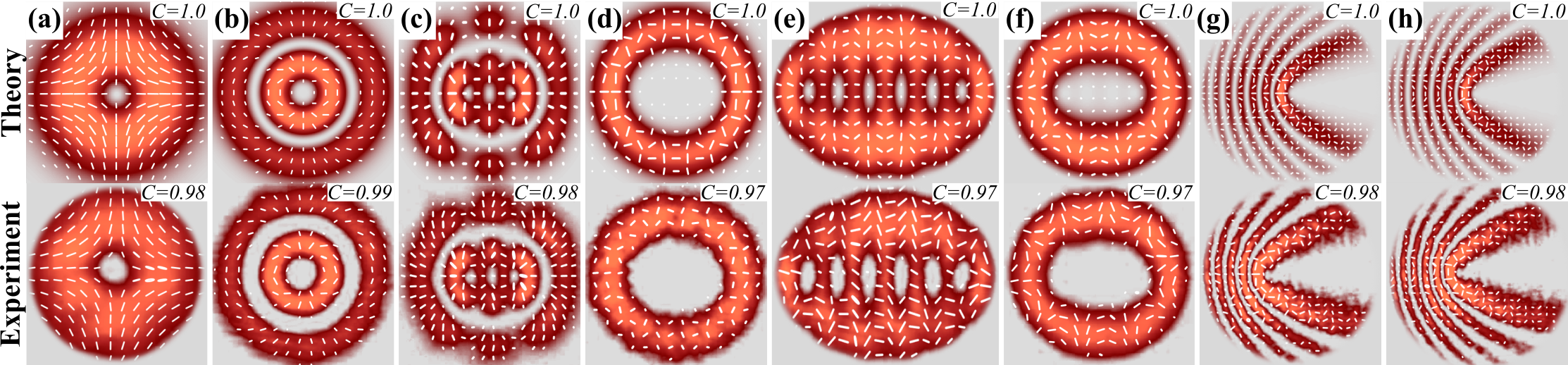}
     \centering
      \caption{Transverse polarisation overlapped with the corresponding intensity profile of exemplary vector modes generated in various coordinate systems.  (a)-(b) $\vect{U}^{\ell,p}_{LG}$;  (c)-(d) $\vect{U}_{IG^h}^{p,m,\varepsilon}$; (e)-(f) $\vect{U}_{MG^h}^{m,e}$, and (f)-(g) parabolic-Gauss vector modes. The specific parameters for each case are detailed in the main text.}
    \label{full_modes}
\end{figure*}

To exemplify the capabilities of our technique, first, we present a comparison between experimental results and numerical simulations of cylindrical vector vortex beams.  Our setup allows a delicate control over the degree of concurrence and polarisation distribution, by digitally adjusting the parameters $\alpha$ and $\theta$.  In figure \ref{Poincare_Sphere} we show the results for the specific case $\vect{U}_{LG}^{1,0}$.  Fig.\ \ref{Poincare_Sphere} (a) shows the corresponding Higher--Order Poincar\'e Sphere (HOPS) with the modes $LG_{0}^{1}$ and $LG_{0}^{-1}$ represented on the North and South poles, respectively.  In such representation, all vector modes with varying degrees of concurrence and inter-modal phases are represented as points $(2\theta, 2\alpha)$ on the surface of the HOPS.  The numbers from 1 to 4 show the location of four cases of pure vector modes with coordinates $(\theta,\alpha)=(\frac{\pi}{4},0), (\frac{\pi}{4},\frac{\pi}{4}), (\frac{\pi}{4},\frac{\pi}{2})$ and $(\frac{\pi}{4},\frac{3\pi}{4})$, this can be understood as a transit along the equator of the HOPS.  The transverse polarisation distribution overlapped with their corresponding intensity profile are shown in Fig.\ \ref{Poincare_Sphere} (b).  In a similar way, the numbers from 5 to 8 correspond to the four vector modes shown in Fig. \ref{Poincare_Sphere} (c) with coordinates $(\theta,\alpha)=(0,0),(\frac{3\pi}{5},0),(\frac{2\pi}{5},0)$ and $(\frac{\pi}{2},0)$.  In this case, the modes follow the path along a geodesic connecting the North and South poles.  Notice the high similarity between the experimental and numerically simulated results.

As a second example, we demonstrate that our experimental setup can generate vector modes from different paraxial families.  The transverse polarisation distribution overlapped with the intensity profile of a representative set of vector modes, generated in different coordinate systems is shown in Fig. \ref{full_modes}. Figures \ref{full_modes} (a) and \ref{full_modes} (b) show  two cylindrical vector modes $\vect{U}_{LG}^{-1,0}$ and $\vect{U}_{LG}^{2,1}$, respectively.  In Fig. \ref{full_modes} (c) and \ref{full_modes} (d) we show two helical Ince-Gauss vector modes $\vect{U}_{IG^h}^{p,m,\varepsilon}$, where the indices $\left \{p, m  \right \}$ $\in$ $\mathbb{N}$ follow the relation  $0 \leq m \leq p$ for even functions and $1 \leq m \leq p$ for odd functions, while $\varepsilon$ represents the eccentricity.  In this case, the spatial mode of each component in Eq.\ \ref{vector_bea}, takes the form of the helical Ince--Gauss modes with opposite handedness, as described in \cite{Liyao2020}.  The parameters $\left \{p,m, \varepsilon  \right \}$ for this specific instance are $\left \{5, 3, 2 \right \}$ and $\left \{8, 8, 2 \right \}$, respectively, and the intermodal phase and weighting parameter are $\left (\alpha, \theta  \right )$ =  $\left(0, \frac{\pi}{4} \right)$. In a similar way, in Fig.\ \ref{full_modes} (e) and \ref{full_modes} (f) we show two helical Mathieu-Gauss vector modes $\vect{U}_{MG^h}^{m,e}$, generated as outlined in \cite{Rosales2021}. The parameters we used for the first case are $a=0.56$, $k_t=16.5\times10^3$ m$^{-1}$, $e=0.9$ and $m=6$, whereas for the second case they are $a=0.2$, $k_t=15.5\times10^3$ m$^{-1}$, $e=0.9$ and $m=4$.  For this mode, $a$ is the semi--minor axis of the elliptical coordinates, $k_t$ is the transverse component of the wave vector, $e$ is the eccentricity of the elliptical coordinates related to the semi-focal distance $f$ as $e=f/a$, and $m$ is the order of the vector mode.

\begin{figure*}[tb]
    \centering 
\includegraphics[width=0.98\textwidth]{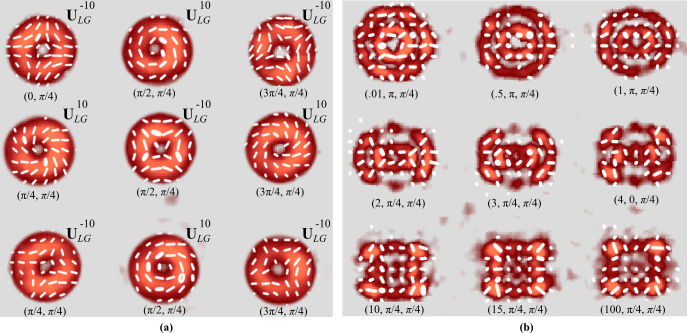}
      \caption{Transverse polarisation distribution overlapped with the corresponding intensity profile of nine multiplexed (a) $\vect{U}^{\ell,p}_{LG}$ and (b) $\vect{U}_{IG^h}^{p,m,\varepsilon}$ vector modes with increasing ellipticity. In (a) the values $p$, $\ell$ are shown in the top-right corner of each mode and the values ($\alpha$, $\theta$) in the bottom of each mode. In (b) the values $p$, $m$ for all the modes where chosen as $p=5$ and $m=3$, respectively, in a similar way, the values ($\varepsilon$,$\alpha$, $\theta$) are shown on the bottom of each mode. Notice the evolution from polar to Cartesian geometry for increasing values of $\varepsilon$.}
    \label{simultaneous_modes}
\end{figure*}

Finally, in Fig. \ref{full_modes} (g) and \ref{full_modes} (h) we show the results for two travelling parabolic-Gauss vector modes $\vect{U}_{TPG}^n$.  In  the first case $\left (\alpha, \theta  \right) = \left (0, \frac{\pi}{4} \right)$, and for the second case   $\left (\alpha, \theta  \right) =  \left (\frac{\pi}{2}, \frac{\pi}{4} \right)$, that where generated with the parameters $n=3$ and $k_t = 1.9\times10^5$m$^{-1}$, for more details please refer to \cite{Hu2021}. Notice the high similarity between the experimental and numerical simulations, which demonstrates qualitatively the high accuracy of this device. In addition, we measured the value of concurrence $C$ for each mode.  The latter quantifies the accuracy of our device.  Finally and with the aim of testing the multiplexing capabilities of our setup, we generated multiple vector beams with independent polarisation distributions and/spatial shape, using a single hologram.  In Fig \ref{simultaneous_modes} (a) we show a representative set of nine different cylindrical vector modes $\vect{U}_{LG}^{\ell,p}$, whose values $\ell$ and $p$ are given as insets in the top-right corner of each mode. In addition, their specific values  $(\alpha, \theta)$ are given below each mode In Fig.\ \ref{simultaneous_modes} (b) we show a set of nine Ince--Gauss vector modes $\vect{U}_{IG^h}^{p,m,\varepsilon}$ which evolve from polar to Cartesian symmetry, by taking different values of eccentricity $\varepsilon$. Their specific values $(\varepsilon ,\alpha, \theta)$ are also given in the bottom of each mode. As a technical note, the generation of the nine simultaneous vector modes implies the use of different values for the linear phase grating by adjusting $\nu, \eta$ in Eq.\ \ref{transmittance_function}, for each mode.  The nature of the DMD and the binary hologram programmed on the modulator, generate a vast amount of diffraction orders.  The components of the desired set of vector modes, propagate along a common optical path and are recombined using a Sagnac interferometer, as described in the experimental setup (Fig.\ \ref{setup_experiment}). Crucially, the background noise of the generated modes depends on the chosen spatial mode basis.  For instance, the noise is higher when using the HIG than for LG spatial basis.  This is due to the spatial filtering, since the generated modes are difficult to isolate from each other, and the diffraction orders may be close to the desired vector modes.  In addition, if the vector modes are separated to reduce the noise, the spatial intensity distribution is degraded.

\section{Conclusions}
In this manuscript we proposed a novel technique to generate arbitrary complex vector modes.  This technique is based on a polarisation insensitive DMD which is becoming very popular in the generation of vector modes as it is a low-cost device, polarisation insensitive and with high refresh rates.  It is precisely its polarisation-insensitive property that is fully exploited in this proposal, which in combination with a common-path interferometer, makes it highly stable.  In this technique, two diagonally polarised, orthogonal modes propagating at different angles are generated simultaneously from a single hologram displayed on the DMD screen.  Both modes enter a triangular Sagnac interferometer, which is formed by a PBS and two mirrors.  Each of the two modes is split into its horizontal and vertical polarisation components by the PBS, and after a round trip exit through a contiguous port.  The two mirrors are adjusted to ensure the overlap between two of the modes with orthogonal polarisation, which generate the vector beam.  We demonstrated this technique experimentally, first by generating arbitrary vector modes on a Higher-order Poincar\'e sphere;  later by generating vector modes with arbitrary transverse profiles,  and finally by generating several vector modes from a single hologram.  This technique might be of high relevance in optical trapping techniques, which require fast reconfiguration of the vector beams \cite{bhebhe2018}.

 
\section*{Funding}
BPG acknowledges support from Consejo Nacional de Ciencia y Tecnolog\'ia (PN2016-3140).  CRG acknowledge support from the National Nature Science Foundation of China (NSFC) under Grant No. 61975047

\section*{Disclosures}
The authors declare that there are no conflicts of interest related to this article.

\section*{References}
\bibliographystyle{iopart-num}

\begin{thebibliography}{10}
\expandafter\ifx\csname url\endcsname\relax
  \def\url#1{{\tt #1}}\fi
\expandafter\ifx\csname urlprefix\endcsname\relax\def\urlprefix{URL }\fi
\providecommand{\eprint}[2][]{\url{#2}}

\bibitem{roadmap}
Rubinsztein-Dunlop H, Forbes A, Berry M~V, Dennis M~R, Andrews D~L, Mansuripur
  M, Denz C, Alpmann C, Banzer P and Bauer T 2017 {\em J. Opt.\/} {\bf 19}
  013001

\bibitem{rosales2018review}
Rosales-Guzm\'{a}n C, Ndagano B and Forbes A 2018 {\em J. Opt.\/} {\bf 20}
  123001

\bibitem{Zhan2009}
Zhan Q 2009 {\em Adv. Opt. Photonics\/} {\bf 1} 1--57

\bibitem{hu2019situ}
Hu X~B, Zhao B, Zhu Z~H, Gao W and Rosales-Guzm{\'a}n C 2019 {\em Optics
  Letters\/} {\bf 44} 3070--3073

\bibitem{toppel2014classical}
T{\"o}ppel F, Aiello A, Marquardt C, Giacobino E and Leuchs G 2014 {\em New
  Journal of Physics\/} {\bf 16} 073019

\bibitem{li2016high}
Li P, Wang B and Zhang X 2016 {\em Optics Express\/} {\bf 24} 15143--15159

\bibitem{Yuanjietweezers2021}
Yang Y, Ren Y, Chen M, Arita Y and Rosales-Guzmán C 2021 {\em Adv.
  Photonics\/} {\bf 3}

\bibitem{Ndagano2017}
Ndagano B, Perez-Garcia B, Roux F~S, McLaren M, Rosales-Guzm\'{a}n C, Zhang Y,
  Mouane O, Hernandez-Aranda R~I, Konrad T and Forbes A 2017 {\em Nature
  Phys.\/} {\bf 13} 397--402

\bibitem{bhebhe2018}
Bhebhe N, Williams P~A~C, Rosales-Guzm{\'a}n C, Rodriguez-Fajardo V and Forbes
  A 2018 {\em Sci. Rep.\/} {\bf 8} 17387

\bibitem{zhao2015high}
Zhao Y and Wang J 2015 {\em Optics Letters\/} {\bf 40} 4843--4846

\bibitem{McLaren2015}
McLaren M, Konrad T and Forbes A 2015 {\em Phys. Rev. A\/} {\bf 92} 023833

\bibitem{Ndagano2016}
Ndagano B, Sroor H, McLaren M, Rosales-Guzm{\'{a}}n C and Forbes A 2016 {\em
  Opt. Lett.\/} {\bf 41} 3407

\bibitem{Zhaobo2020}
Zhao B, Hu X~B, Rodr\'iguez-Fajardo V, Forbes A, Gao W, Zhu Z~H and
  Rosales-Guzm\'an C 2020 {\em Appl. Phys. Lett.\/} {\bf 116} 091101

\bibitem{Aiello2015}
Aiello A, T{\"o}ppel F, Marquardt C, Giacobino E and Leuchs G 2015 {\em New J.
  Phys.\/} {\bf 17} 043024

\bibitem{konrad2019quantum}
Konrad T and Forbes A 2019 {\em Contemporary Physics\/}  1--22

\bibitem{Qian2015}
Qian X~F, Little B, Howell J~C and Eberly J~H 2015 {\em Optica\/} {\bf 2}
  611--615

\bibitem{Bhebhe2018a}
Bhebhe N, Rosales-Guzman C and Forbes A 2018 {\em Appl. Opt.\/} {\bf 57}
  5451--5458

\bibitem{Maurer2007}
Maurer C, Jesacher A, F{\"{u}}rhapter S, Bernet S and Ritsch-Marte M 2007 {\em
  New J. Phys.\/} {\bf 9} 78

\bibitem{Rosales2017}
Rosales-Guzm\'{a}n C, Bhebhe N and Forbes A 2017 {\em Opt. Express\/} {\bf 25}
  25697--25706

\bibitem{Moreno2012}
Moreno I, Davis J~A, Hernandez T~M, Cottrell D~M and Sand D 2012 {\em Opt.
  Express\/} {\bf 20} 364--376

\bibitem{Mitchell2017}
Mitchell K~J, Radwell N, Franke-Arnold S, Padgett M~J and Phillips D~B 2017
  {\em Opt. Express\/} {\bf 25} 25079--25089

\bibitem{Rong2014}
Rong Z~Y, Han Y~J, Wang S~Z and Guo C~S 2014 {\em Opt. Express\/} {\bf 22} 1636

\bibitem{SPIEbook}
Rosales-Guzm\'{a}n C and Forbes A 2017 {\em How to shape light with spatial
  light modulators\/} SPIE.SPOTLIGHT (SPIE Press)

\bibitem{Hu2018}
Hu X, Zhao Q, Yu P, Li X, Wang Z, Li Y and Gong L 2018 {\em Opt. Express\/}
  {\bf 26} 1796--1808

\bibitem{Ren2015}
Ren Y~X, Lu R~D and Gong L 2015 {\em Annalen der Physik\/} {\bf 527} 447--470

\bibitem{Chen2015DMD}
Chen Y, Fang Z~X, Ren Y~X, Gong L and Lu R~D 2015 {\em Appl. Opt.\/} {\bf 54}
  8030--8035

\bibitem{Mitchell2016}
Mitchell K~J, Turtaev S, Padgett M~J, \v{C}i\v{z}m\'{a}r T and Phillips D~B
  2016 {\em Opt. Express\/} {\bf 24} 29269--29282

\bibitem{Goorden2014}
Goorden S~A, Bertolotti J and Mosk A~P 2014 {\em Opt. Express\/} {\bf 22}
  17999--18009

\bibitem{Lerner2012}
Lerner V, Shwa D, Drori Y and Katz N 2012 {\em Opt. Lett.\/} {\bf 37}
  4826--4828

\bibitem{Hu2021Random}
Hu X~B, Ma S~Y and Rosales-Guzm{\'{a}}n C 2021 {\em J. Opt.\/} {\bf 23} 044002

\bibitem{Scholes2019}
Scholes S, Kara R, Pinnell J, Rodr{\'\i}guez-Fajardo V and Forbes A 2019 {\em
  Optical Engineering\/} {\bf 59} 1 -- 12

\bibitem{Hu2022}
Hu X~B and Rosales-Guzm{\'{a}}n C 2022 {\em Journal of Optics\/} {\bf 24}
  034001

\bibitem{Rosales2020}
Rosales-Guzm{\'a}n C, Hu X~B, Selyem A, Moreno-Acosta P, Franke-Arnold S,
  Ramos-Garcia R and Forbes A 2020 {\em Scientific Reports\/} {\bf 10} 10434

\bibitem{Gong2014}
Gong L, Ren Y, Liu W, Wang M, Zhong M, Wang Z and Li Y 2014 {\em J. Appl.
  Phys.\/} {\bf 116} 183105

\bibitem{Liyao2020}
Yao-Li, Hu X~B, Perez-Garcia B, Bo-Zhao, Gao W, Zhu Z~H and Rosales-Guzm{\'a}n
  C 2020 {\em Applied Physics Letters\/} {\bf 116} 221105

\bibitem{Zhao2022}
Zhao B, Rodríguez-Fajardo V, Hu X~B, Hernandez-Aranda R~I, Perez-Garcia B and
  Rosales-Guzmán C 2022 {\em Nanophotonics\/} {\bf 11} 681--688

\bibitem{Rosales2021}
Rosales-Guzmán C, Hu X, Rodríguez-Fajardo V, Hernandez-Aranda R~I, Forbes A
  and Perez-Garcia B 2021 {\em J. Opt.\/} {\bf 23} 034004

\bibitem{Hu2021}
Hu X~B, Perez-Garcia B, Rodr\'{i}guez-Fajardo V, Hernandez-Aranda R~I, Forbes A
  and Rosales-Guzm\'{a}n C 2021 {\em Photon. Res.\/} {\bf 9} 439--445

\bibitem{hu2021generation}
Hu X and Rosales-Guzm{\'a}n C 2021 {\em Journal of Optics\/}

\bibitem{Perez-Garcia2017}
Perez-Garcia B, L\'{o}pez-Mariscal C, Hernandez-Aranda R~I and
  Guti\'{e}rrez-Vega J~C 2017 {\em Appl. Opt.\/} {\bf 56} 6967--6972

\bibitem{Lee79}
Lee W~H 1979 {\em Appl. Opt.\/} {\bf 18} 3661--3669

\bibitem{Galvez2015}
Galvez E~J 2015 {\em Light Beams with Spatially Variable Polarization\/}
  (Wiley-Blackwell) chap~3, pp 61--76 ISBN 9781119009719

\bibitem{galvez2012poincare}
Galvez E~J, Khadka S, Schubert W~H and Nomoto S 2012 {\em Applied optics\/}
  {\bf 51} 2925--2934

\bibitem{hariharan2003optical}
Hariharan P 2003 {\em Optical Interferometry, 2e\/} (Elsevier)

\bibitem{Goldstein2011}
Goldstein D~H 2011 {\em Polarized light\/} (CRC Press)

\bibitem{Selyem2019}
Selyem A, Rosales-Guzm\'an C, Croke S, Forbes A and Franke-Arnold S 2019 {\em
  Phys. Rev. A\/} {\bf 100}(6) 063842

\bibitem{Zhao2019}
Zhao B, Hu X~B, Rodr\'{i}guez-Fajardo V, Zhu Z~H, Gao W, Forbes A and
  Rosales-Guzm\'{a}n C 2019 {\em Opt. Express\/} {\bf 27} 31087--31093

\bibitem{Forbes2014book}
Forbes A 2014 {\em Laser Beam Propagation: Generation and Propagation of
  Customized Light\/} (Taylor \& Francis) ISBN 9781466554399

\end{thebibliography}
\providecommand{\newblock}{}

\end{document}